# Mesh of Spatiotemporal Optical Vortices with Programmable Intensity Nulls

Jinxin Wu, Dan Wang, Qingqing Liang, Jianhua Hu, Jiahao Dong, Jijun Feng, Yi Liu

*Shanghai Key Lab of Modern Optical System, University of Shanghai for Science and Technology, 516, Jungong Road, 200093 Shanghai, China*

**ABSTRACT:** Light carrying transverse orbital angular momentum (T-OAM) in the form of spatiotemporal optical vortices (STOVs) is opening new degrees of freedom for structured light manipulation. Such spatiotemporal wavepackets hold significant potential for optical trapping, analog optical computing, studying photonic symmetry and topology, among others. Up to now, synthesizing of such vortices is limited in one dimension, either in temporal or spatial domain. In this work, we propose and experimentally demonstrate a two-dimensional flexible mesh of spatiotemporal optical vortices (M-STOV) with programmable intensity nulls, and analyze their diffraction patterns for detection. Furthermore, we extend the spectral range of M-STOV via second-harmonic generation while examining the transfer of OAM in this nonlinear process. This study establishes a foundational framework for designing higher dimensional spatiotemporal vortex fields and promises a high-capacity information carrier based on ST optical vortices.

## INTRODUCTION

The angular momentum of light comprises both spin angular momentum (SAM) and orbital angular momentum (OAM), which are fundamental properties of light with broad applications across numerous fields. While SAM is associated with light polarization, OAM of light is related to phase vortices (*1*). Optical OAM is generally quantized by an integer quantum number *l*, also known as the topological charge. Based on its direction relative to the wave propagation vector, optical OAM can be further classified as longitudinal OAM (L-OAM), which aligns with the wave vector in paraxial beams, and transverse OAM (T-OAM), whose orientation is perpendicular to the wave vector. First identified by Allen et al. in 1992 (*2*), L-OAM has been widely exploited in optical communications (*3-5*), optical tweezers (*6-8*), holography (*9, 10*), optical microcomb (*11, 12*) and other domains. More recently, transverse OAM (T-OAM), manifested as spatiotemporal optical vortices (STOVs) as illustrated in Fig. 1(a), was initially observed from nonlinear filamentary propagation of femtosecond intense pulse in air by Jhajj et al. (*13*). Subsequent developments enabled the linear generation of STOVs (*14, 15*), stimulating extensive research into their propagation and diffraction (*16-18*), detection (*19-21*), reflection and refraction (*22*), nonlinear phenomena (*23-26*) as well as advanced topological structures based on STOVs (*27-29*).

Among these investigations, a notable research focus is the manipulation of fields with multiple phase singularities, in order to expand the capacity of information transfer based on optical vortices.

Experimental realizations of first-order time-varying phase singularities (*30*) laid the foundation for extending this concept to spatially varying multi-singularity structures, in terms of STOV strings, which show promising potential for optical communications (*31*). Up to now, realization and manipulation of higher-dimensional spatiotemporal singularity structures have remained underexplored.

In this work, we develop a 2-dimension transverse OAM structure, coined as Mesh of Spatiotemporal Optical Vortices with Programmable Intensity Nulls (M-STOV), through integrated experimental and theoretical approaches. As depicted in Fig. 1(b), each phase singularity acts analogously to a "PIN" embedded within a Gaussian envelope, creating intensity nulls in the spatiotemporal domain. As interference-based methods struggle to characterize STOVs with diverse T-OAM modes due to their rapidly varying spatiotemporal phases (*31*), we employ a diffraction-based detection approach (*18*). As a result of this method, the space-spectrum of M-STOV exhibits intricate structural features, which can be understood as the comprehensive effect arising from all individual singularities. Furthermore, we extend the spectral range of M-STOV from the near-infrared to the ultraviolet regime, via the nonlinear second harmonic generation (SHG). This work expands the conceptual landscape of spatiotemporal optical structures and offers new insights into the fundamental properties of structured light.

**RESULT**

**Experiment Setup**

We employed a folded 4f pulse shaper (Fig. 2) (*25*) to generate M-STOV, which consist of a diffraction grating, cylindrical lens, and spatial light modulator (SLM). In the experiments, we utilized a mode-locked laser source centered at 800 nm, which was collimated, beam-expanded, and then incident onto the grating-cylindrical lens system for spectrum-space mapping. By loading designed phase patterns on the SLM plane, we achieved spatiotemporal modulation of the pulses. The shaped pulses were subsequently focused by a spherical lens to form M-STOV around its focal plane. The diffraction patterns of M-STOV are formed by a grating-cylindrical lens system and recorded using a charge-coupled device camera (CCD). For nonlinear frequency conversion of the structured light field, after the spherical lens the M-STOV passed through a beta barium borate (BBO) crystal.

The phase patterns loaded on the SLM can be mathematically described by:

$$\varphi = \sum_{j=1}^{n} exp\left(i sgn(l_j) tan^{-1}\left(\frac{y+d_{yj}}{x+d_{xj}}\right)\right) \tag{1}$$

where $x$ and $y$ correspond to the spatial coordinates of the SLM plane, respectively. The parameters $d_{xj}$ and $d_{yj}$ denotes the displacement between the j-th STOV and reference point of the SLM along x and y directions, respectively. The integer topological charge $l_j$ determines the vortex order, with $sgn(l_j)$ accounting for the helicity of each phase singularity. In Fig. 2, we present the phase pattern designed to generate an M-STOV structure with $\begin{bmatrix} 1 & 1 \\ 1 & 1 \end{bmatrix}$ singularities distribution. The relationship between singularity displacements in the frequency-spatial domain and those in the time-space domain is presented in Supplementary Material.

**The formulation, generation and detection of M-STOV**

Under the scalar, paraxial, slowly varying envelope, and narrow bandwidth approximations, the electric field of an optical wave packet propagating in a uniform medium with anomalous group velocity dispersion (GVD) can be expressed as the solution of (*29, 32*):

$$\frac{\partial^2 \psi}{\partial x^2} + \frac{\partial^2 \psi}{\partial y^2} + \frac{\partial^2 \psi}{\partial t^2} + 2i\frac{\partial \psi}{\partial z} = 0 \qquad (2)$$

where $\psi$ represents the scalar wavefunction, $x$ and $y$ denote the normalized transverse coordinates, $t$ is the normalized retarded time, and $z$ stands for the propagation distance. Given that $\psi$ varies slowly in the *x*-direction in our current setup, its second derivative $\frac{\partial^2 \psi}{\partial x^2}$ can be neglected.

As a solution of Eq. (2) in the Cartesian coordinates (y, t), M-STOV at $z=0$ can be expressed as:

$$\psi(y,t) \propto \prod_{j=1}^{n} \left(\frac{t-t_j}{w_{tj}} + i\,sgn(l_j)\frac{y-y_j}{w_{yj}}\right)^{|l_j|} exp\left(-\frac{y^2}{w_y^2} - \frac{t^2}{w_t^2}\right) \qquad (3)$$

where $w_y$ and $w_t$ indicate the spatial width and the temporal width of the Gaussian envelope individually. $l_j$ is integer that denotes topological charge of each spatiotemporal singularity. $y_j$ and $t_j$ are the displacement in space and time of j-th singularity, respectively. $w_{yj}$ and $w_{tj}$ mean the spatial width and the temporal width of j-th singularity.

In the detection system, the optical field distribution on the CCD plane formed after the beam propagates through the grating-cylindrical lens system, represents the temporal Fourier transform of the incident optical field (*33, 34*). Spectrum of M-STOV can be calculated as:

$$\mathcal{F}_t[\psi(y,t)] \qquad (4)$$

Rows 1 and 2 of Fig. 3 present the simulated spatiotemporal intensity and phase distribution for the mesh of singularities distribution with $M = \begin{bmatrix} 0 & 0 \\ 1 & 0 \end{bmatrix}$, $\begin{bmatrix} 1 & 0 \\ 0 & 1 \end{bmatrix}$, $\begin{bmatrix} 1 & 0 \\ 1 & 1 \end{bmatrix}$ and $\begin{bmatrix} 1 & 1 \\ 1 & 1 \end{bmatrix}$ generated from the 4f shaper. The first row presents the spatiotemporal intensity distributions of the M-STOV, with the second row illustrating their characteristic spiral phases. While a conventional STOV has a uniform energy distribution around its central singularity, a M-STOV exhibits a highly irregular profile determined by its pre-engineered singularity mesh. Each singularity embodies a $2\pi$ phase winding. Consequently, the total topological charge is defined as the sum of these windings.

In rows 3 and 4 of Fig. 3, we show the simulated and experimental detection results of M-STOV, where the displacements for each singularity are $d_{yj} = 0.5mm$ and $d_{xj} = 1.25mm$. The third row shows the simulated space-spectra of the M-STOV generated by the 4f system, and the fourth row presents the corresponding experimentally results. The diffraction patterns of M-STOV feature multi-lobe structures, where the number of lobes correlates with the total topological charges. Each gap between lobes corresponds to one unit of topological charge, or a $2\pi$ phase winding, in a manner analogous to conventional STOVs of the same total charge (*18*). However, a key distinction arises from the displacement of the singularities: unlike conventional STOVs which exhibit a spectral energy distribution characterized by "strong side lobes and a weak center," M-STOV often have their spectral peak shifted away from these side lobes.

To further demonstrate flexibility of the topological charge order and position of singularities in the spatiotemporal domain, we present two representative cases in Fig. 4. The first is a generalized M-STOV with irregularly arranged singularities, presented in the first row. The other is an M-STOV with higher topological charges $M = \begin{bmatrix} 2 & 0 \\ 0 & -1 \end{bmatrix}$, shown in the second row. In Fig. 4, the white texts indicate the topological charge of the phase singularities. M-STOV comprising topological charges of different helicities forms intricate diffraction patterns, with a detailed analysis of these patterns is provided in the section below.

**Analysis of the diffraction detection of M-STOV**

Now we turn to the analysis of detection of the M-STOV. Dividing Eq (4) into two parts and set:

$$f_{j,j \neq a}(y, \omega) = \mathcal{F}_t \left[ \left( \frac{t-t_j}{w_{tj}} + i\, sgn(l_j) \frac{y-y_j}{w_{yj}} \right)^{|l_j|} \right] \tag{5}$$

$$g(y, \omega) = \mathcal{F}_t \left[ \left( \frac{t-t_a}{w_{ta}} + i\, sgn(l_a) \frac{y-y_a}{w_{ya}} \right)^{|l_a|} exp\left( -\frac{y^2}{w_y^2} - \frac{t^2}{w_t^2} \right) \right] \tag{6}$$

The first part, $f_{j,j \neq a}(y, \omega)$, corresponds to intensity nulls (excluding the $a$-th one) embedded in the pulse, whereas the second part, $g(y, \omega)$, describes a Gaussian envelope containing a single intensity null with both temporal and spatial displacements $t_a$ and $y_a$. Calculate Eq (5) and Eq (6), respectively. We find that $f_{j,j \neq a}(y, \omega)$ comprises a series of derivatives of Dirac delta functions ($\sum \delta^k(\omega)$), while the final form of $g(y, \omega)$ is described by a tilt $|l_a|$-order Hermite-Gaussian distribution. According to the convolution theorem, Eq (4) can be regarded as the convolution of $f_{j,j \neq a}(y, \omega)$ and $g(y, \omega)$. As the convolution between the k-th derivative of the Dirac delta function and an arbitrary function yields the k-th derivative of that function evaluated at zero (*35*), and the Hermite polynomials maintain their functional form under arbitrary-order differentiation (*36*), the intensity pattern on the CCD can be viewed as the combination of tilted Hermite-Gaussian distributions corresponding to the topological charge orders $|l_j|$.

We noticed that the optical space-spectrum of a multi-singularities field is not a simple superposition of individual sub-spectra. Due to the finite interaction range of each singularity, their collective effect (convolution) effectively "averages" the spectrum of a single singularity, leading to broadening and smoothing of the overall spectral features. Therefore, the final spectrum retains some characteristics of the individual sub-spectra but does not constitute a simple superposition of them. This analysis aligns with our experimental phenomena. The complete spectral analytical expressions for the two singularities in our M-STOV, along with detailed derivations of all calculations, are provided in the Supplementary Material.

To illustrate this combination effect of the space-spectrum for detection of M-STOV, in our experiment, we generate a M-STOV with the topological charges arrange $M = \begin{bmatrix} -1 & -1 \\ 1 & -1 \end{bmatrix}$ to analyze its space-spectrum, with singularity displacements $d_{yj} = 0.5mm$ and $d_{xj} = 1.25mm$ introduced on the SLM. The resulting space-spectrum demonstrates a superposition of multiple sub-spectra, as presented in Fig. 5. As a representative case, we can decompose the topological charges arrange as $M = M_1 \boxplus M_2$, with $M_1 = \begin{bmatrix} 0 & -1 \\ 0 & -1 \end{bmatrix}$ and $M_2 = \begin{bmatrix} -1 & 0 \\ 1 & 0 \end{bmatrix}$. Let us consider the following configuration:

$$\begin{bmatrix} 0 & -1 \\ 0 & -1 \end{bmatrix} \boxplus \begin{bmatrix} -1 & 0 \\ 1 & 0 \end{bmatrix} = \begin{bmatrix} -1 & -1 \\ 1 & -1 \end{bmatrix}$$

The white texts in Fig. 5 indicate the topological charge of the phase singularities. The boxed plus symbol ($\boxplus$) is introduced here to distinguish this specific process from ordinary superposition. Additional combination cases are provided in the Supplementary Material.

As shown in the rightmost column of Fig. 5, the detection pattern for element $M = \begin{bmatrix} -1 & -1 \\ 1 & -1 \end{bmatrix}$ resembles a "butterfly". The overall shape of this "butterfly" structure is primarily comprised of the detection pattern for element $M_1 = \begin{bmatrix} 0 & -1 \\ 0 & -1 \end{bmatrix}$ (The first column), while the form of its lower-right wing is modulated by the diffraction pattern of element $M_2 = \begin{bmatrix} -1 & 0 \\ 1 & 0 \end{bmatrix}$ (The second column). The superposition region of elements $M_1 = \begin{bmatrix} 0 & -1 \\ 0 & -1 \end{bmatrix}$ and $M_2 = \begin{bmatrix} -1 & 0 \\ 1 & 0 \end{bmatrix}$ is smoothed by the convolution process, resulting in the distortion of certain distinct.

**Second-harmonic generation of M-STOV**

To expand the available wavelength of the M-STOV and to examine the angular momentum transfer in the nonlinear optical process, the beam was directed through a BBO crystal to produce its second harmonic. Based on the simple formulation of SHG (*37*):

$$E^{(2\omega)}(x,y,\zeta) = E_0 e^{il^{(2\omega)}\phi(x,y,\zeta)} \propto \left(E^{(\omega)}(x,y,\zeta)\right)^2 \tag{7}$$

Accordingly, the SHG field of M-STOV can be written as:

$$E^{(2\omega)} \propto \left(\psi(y,t)\right)^2 \propto \prod_{j=1}^{n} \left(\frac{t-t_j}{w_{tj}} + isgn(l_j)\frac{y-y_j}{w_{yj}}\right)^{|2l_j|} exp\left(-\frac{2y^2}{w_y^2} - \frac{2t^2}{w_t^2}\right) \tag{8}$$

From Eq. (8), it can be inferred that each phase singularity in the SHG field of the M-STOV carries a topological charge that is twice of the original value, while the total topological charge is also doubled.

Our SHG results for the M-STOV are presented in Fig. 6, where the selected parameters are $M = \begin{bmatrix} 1 & 1 \\ 1 & 1 \end{bmatrix}$. Fig. 6(a) shows the detection pattern for the 800 nm field, and Fig. 6(b) displays the optical field spectrum at 800 nm measured with a spectrometer. Fig. 6(c) and (d) present the diffraction pattern and the spectrum of the generated 400nm M-STOV. In Fig. 6(a), there are five bright lobes, indicating a total angular momentum of 4. In Fig. 6(c), nine bright lobes are observed, corresponding to a total angular momentum of 8. This observation confirms the doubling of total angular momentum during second-harmonic generation for M-STOV. We have also exploited other forms of M-STOV and have always observed the doubling of topological charges. Some other results of the SHG field from M-STOV are included in the Supplementary Material.

**DISCUSSION**

In conclusion, we have theoretically formulated and experimentally generated a structured optical vortex characterized by programmable orbital angular momentum (OAM), referred to as the mesh of spatiotemporal optical vortices with programmable intensity nulls (M-STOV), and successfully extended its spectral range via second-harmonic generation. Furthermore, the space-spectrum of this complex two-dimensional spatiotemporal structure was analyzed, thereby extending the applicability of the diffraction-based detection method for STOV characterization. This work broadens the family of spatiotemporally structured light fields and paves way for potential application in many fields, such as optical communication, optical trapping and light-matter interaction. Nonetheless, the characterization of such complicated spatiotemporal optical wavepacket remains challenging. Our current approach, while capable of revealing the topological charge number *l*, does not allow full phase retrieval, highlighting the need for more advanced diagnostic tools in future studies.

## MATERIALS AND METHODS
### Experimental setup and configuration
In our experimental setup, a 4f pulse shaper generated the STOV light fields at fundamental-frequency. The laser source was a commercial femtosecond system (Spectral-Physics SOL-35F-1K-HP-T), producing 35-fs pulses at 800 nm center wavelength with up to 7 mJ pulse energy at 1 kHz repetition rate. As illustrated in Fig. 2, the first grating (G1) dispersed the pulses spatially at 46° incidence, after which a cylindrical lens (CL1; f = 150 mm) focused the beam to accomplish Fourier transformation. A spatial light modulator (SLM; Hamamatsu X13138) placed at the CL1 focal plane imposed a spiral phase profile in the ω–y domain. The SLM-reflected pulse traveled back through CL1, underwent an inverse Fourier transform, and was reflected again by G1 before being directed out through a beam splitter (BS). A subsequent convex lens (f = 1000 mm) executed Fourier transformation to produce the far-field STOV matrix. Type I BBO crystals ($\theta$ = 29.2°; thickness: 0.05 mm) served as the nonlinear medium for second-harmonic generation (SHG) of the linearly polarized STOV beams. Spatiotemporal topological charges were measured via a diffraction-based approach, wherein the beam was incident on a diffraction grating (1200 lines/mm) and focused by a cylindrical lens (CL2; f = 100 mm), with the resulting diffraction patterns captured at the focal plane using a CCD. Wavelength selection was achieved using optical filters, and the distances L1 and L2 between the cylindrical lens and grating/SLM were both set to 150 mm.

**Funding:**
This work was supported by National Natural Science Foundation of China (12034013, U23A20381).

**Author contributions:**
Y.L., and J.F., Q.L. conceived the idea, J.W., J.H., J.D. performed the theoretical derivation and the simulation, J.W., D.W., and Q.L. performed the experimental study, J.W., and Y.L. drafted the manuscript. Y.L., and J.F. supervised the project. All authors approved the final version of the manuscript.

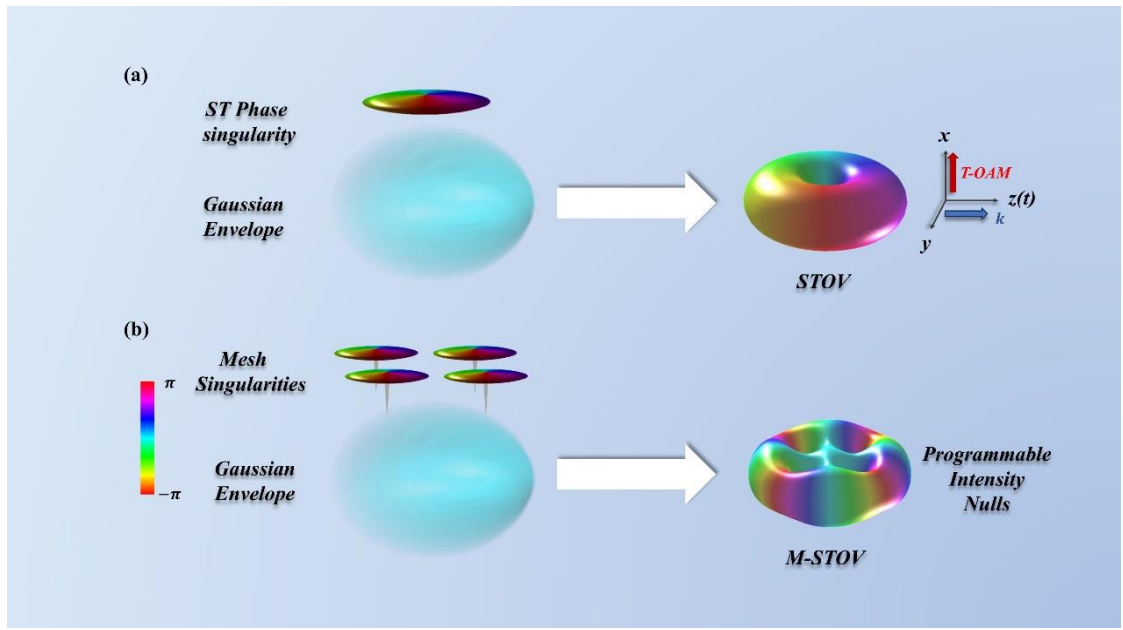

**Fig. 1. Generation of a Spatiotemporal Optical Vortex (STOV) and a Mesh of STOV with Programmable Intensity Nulls (M-STOV). a**, Imprint a spiral phase onto a Gaussian beam generates a spatiotemporal optical vortex (STOV). **b**, Apply programmable intensity nulls to a Gaussian envelope can generate M-STOV.

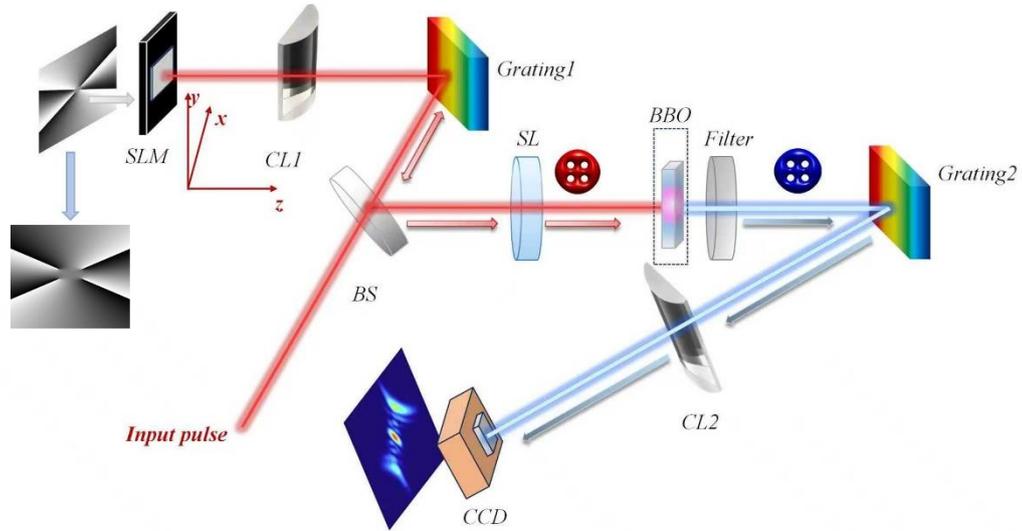

**Fig. 2. Schematic of the experiment setup for generation and detection of M-STOV.** In the experiment, fundamental M-STOV pulses are generated from a folded 4f pulse shaper. Second-harmonic M-STOV pulses are generated in a BBO crystal. The pulses are characterized by a grating-cylindrical lens system, and the detection patterns are recorded by a CCD camera. BS, beam splitter; CL, cylindrical lens; SL, spherical lens; BBO, beta barium borate crystal; SLM, spatial light modulator; CCD, charge coupled device.

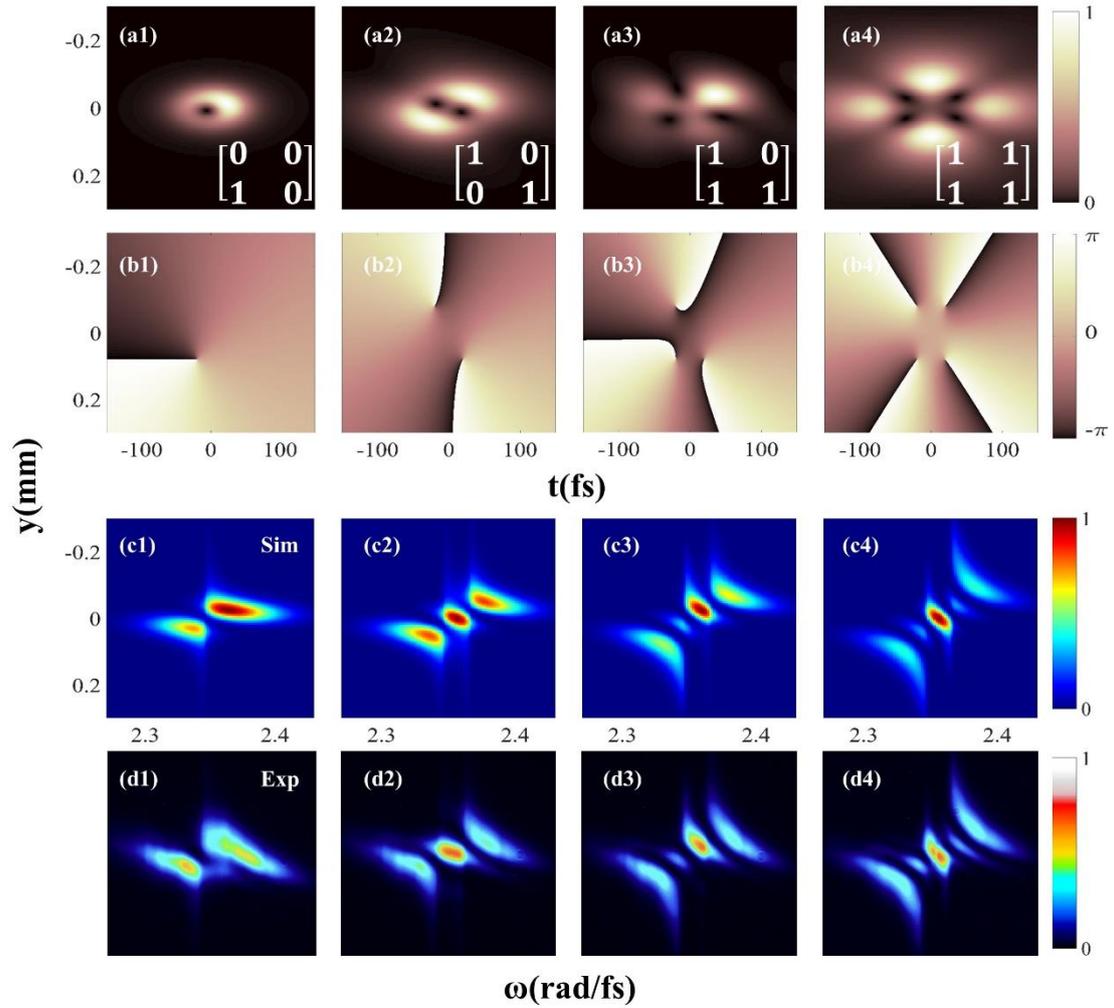

**Fig. 3. Simulated and experimental results of M-STOV.** Row 1 (**a1-a4**) corresponds to the simulated distribution of the M-STOV from the 4f shaper, while row 2 (**b1-b4**) presents the spiral phase in spatiotemporal domain. The third row (**c1-c4**) shows the simulated space-spectrum of the M-STOV generated by a 4f system. The experimental results, recorded by a CCD camera, are displayed in row4 (**d1-d4**). The white text indicates the mesh of topological charges for the simulation and experiments.

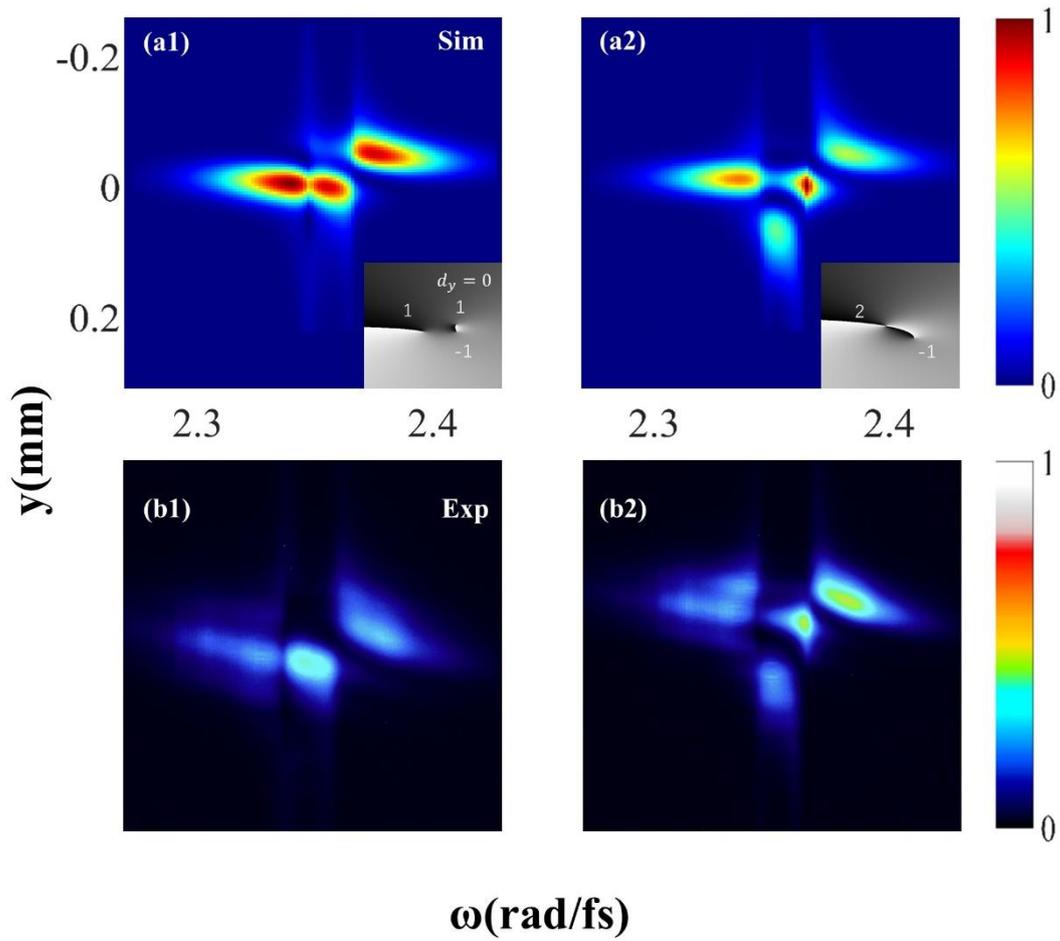

Fig. 4. **Simulated and experimental results of a generalized M-STOV with irregularly arranged singularities and a M-STOV with higher order topological charges.** Row 1 (**a1, a2**) shows the corresponding simulated diffraction patterns. Row 2 (**b1, b2**) presents the measured diffraction patterns. Phase patterns loaded on SLM are inserted at the bottom-right corner of the simulation results. The white text indicates the mesh of topological charges for the simulation and experiments.

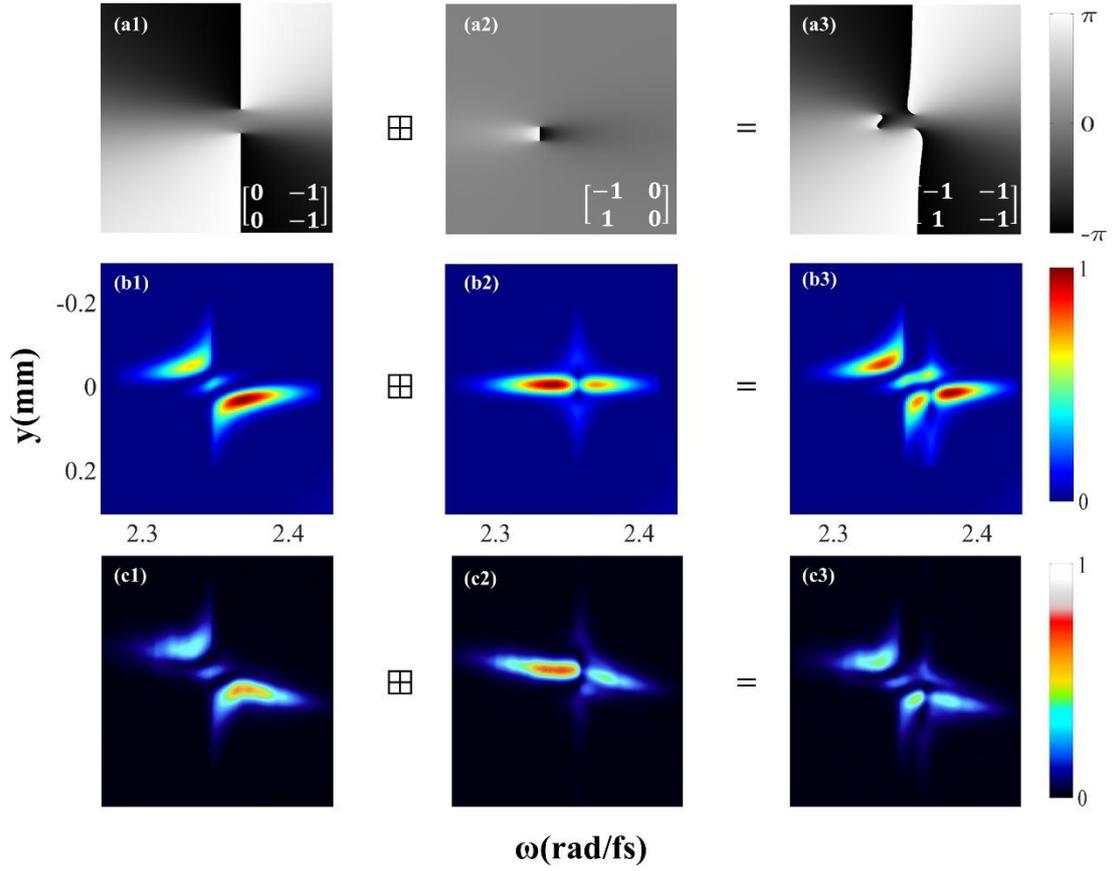

**Fig. 5.** Simulated and experimental results of M-STOV with topological charges arrangement $\begin{bmatrix} 0 & -1 \\ 0 & -1 \end{bmatrix} \boxplus \begin{bmatrix} -1 & 0 \\ 1 & 0 \end{bmatrix} = \begin{bmatrix} -1 & -1 \\ 1 & -1 \end{bmatrix}$. Row 1 (**a1–a3**) displays the composition of phase patterns loaded on SLM. Row 2 (**b1–b3**) illustrates the simulated measurement results. Row 3 (**c1–c3**) contains the corresponding experimental space-spectra. The boxed plus symbol (⊞) is introduced here to distinguish this specific process from ordinary superposition. The white text indicates the mesh of topological charges for the simulation and experiments.

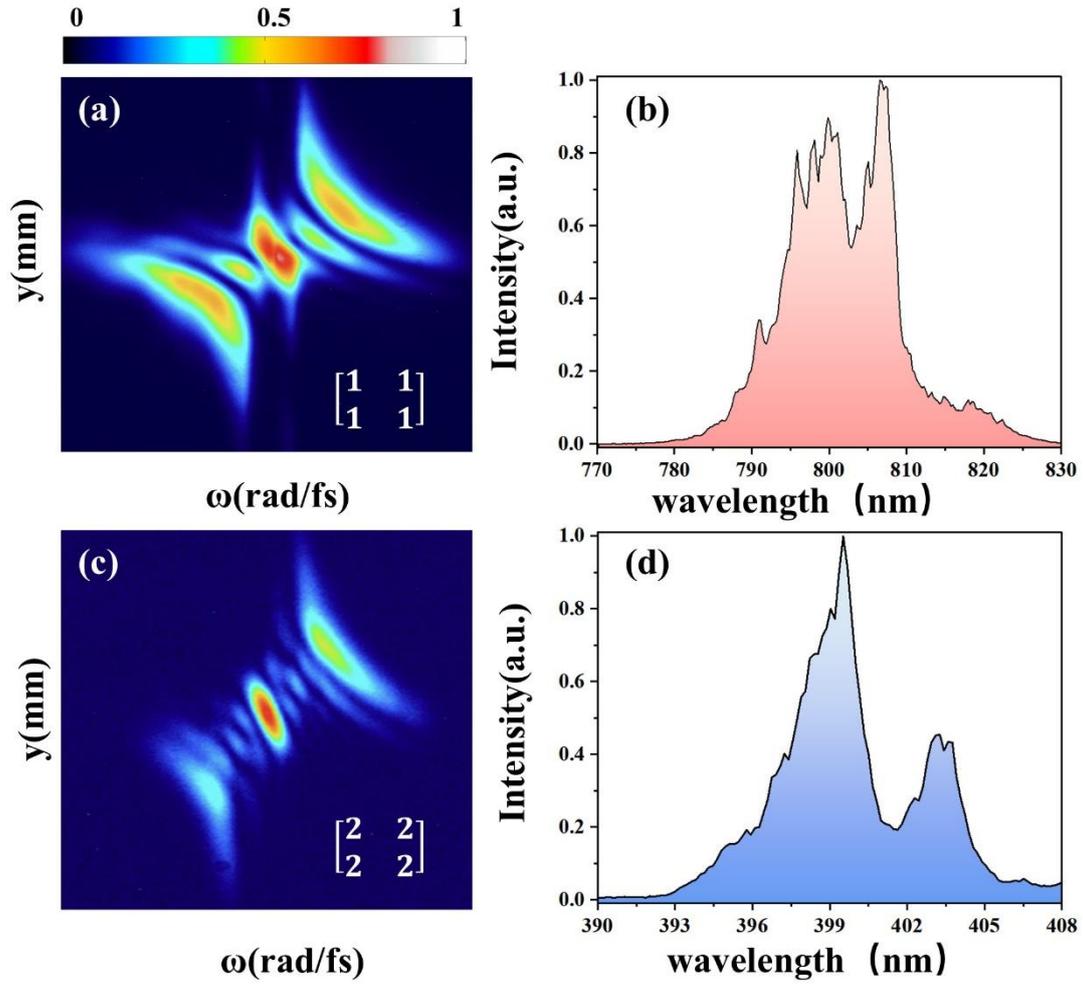

**Fig. 6. Measured space-spectrum of M-STOV with topological charges arrangement $\begin{bmatrix} 1 & 1 \\ 1 & 1 \end{bmatrix}$ at 800nm and 400nm, as well as the related spectrums.** (**a**) and (**b**) the diffraction pattern and the spectrum of 800nm M-STOV. (**c**) and (**d**) the diffraction pattern and the spectrum of 400nm M-STOV.

# Supplementary materials

**The relationship between frequency-domain singularity displacement and time-domain singularity displacement**

Since the inverse Fourier transform of the optical field on the SLM corresponds to a non-integrable expression, we can only investigate the relationship between frequency-domain singularity displacement and time-domain singularity displacement through the spectrum captured by the CCD and numerical simulations.

As the theory depicted in the formulation and spectrum of M-STOV-PIN, the CCD plane is the frequency-spatial plane (y-ω). Hence, the spectrum of general STOV with singularity displacement can be calculated as:

$$\int_{-\infty}^{\infty} \left( \frac{t-t_0}{w_t} + i sgn(l) \frac{y-y_0}{w_y} \right)^{|l|} exp\left( -\frac{t^2}{w_t^2} - \frac{y^2}{w_y^2} \right) e^{-i\omega t} dt \tag{S1}$$

Eq. (S1) reads:

$$A_0 H_{|l|} \left( sgn(l) \frac{(y-y_0)}{w_y} - \frac{\omega w_t}{2} + i \frac{t_0}{w_t} \right) exp\left( -\frac{y^2}{w_y^2} - \frac{w_t^2 \omega^2}{4} \right) \tag{S2}$$

where $A_0$ is a complex function. $H_{|l|}(\cdot)$ is the Hermite polynomial of order $|l|$. $x_0$ is the displacement of the singularity in space, while $t_0$ means the temporal displacement.

Experimental and simulated results of first-order STOV generated by shifting the phase singularity position in both SLM frequency and spatial domains are presented in Fig. S1. Fig. S1(a1) shows the SLM phase pattern without displacement. Fig. S1(b1) presents the simulated spatiotemporal distribution of STOV without displacement. Fig. S1(c1) displays the simulated STOV diffraction pattern without displacement. Fig. S1(d1) shows the measured STOV diffraction pattern without displacement. Fig. S1(a2) illustrates the SLM phase pattern with frequency-domain displacement. Fig. S1(b2) presents the simulated spatiotemporal distribution with frequency displacement. Fig. S1(c2) shows the simulated STOV diffraction pattern with frequency displacement. Fig. S1(d2) displays the measured STOV diffraction pattern with frequency displacement. Fig. S1(a3) depicts the SLM phase pattern with spatial-domain displacement. Fig. S1(b3) shows the simulated spatiotemporal distribution with spatial displacement. Fig. S1(c3) presents the simulated STOV diffraction pattern with spatial displacement. Fig. S1(d3) displays the measured STOV diffraction pattern with spatial displacement. All intensity plots in the figure are normalized for consistency.

The definition of the Hermite polynomial is (*38*):

$$H_n(\xi) = (-1)^n e^{\xi^2} \frac{d^n}{d\xi^n} e^{-\xi^2} = p(\xi) \tag{S3}$$

This is a polynomial in ξ with the highest order $(2\xi)^n$, whose value grows increasingly larger as ξ increases. The Gaussian distribution introduces a decay to the Hermite polynomial as ξ approaches infinity, ultimately balancing the two effects. At this equilibrium, the STOV spectrum with a topological charge of 1 exhibit uniform energy distribution in its two sidelobes. However, introducing a spatial displacement $x_0$ to the singularity introduces an additional real term $sgn(l)\frac{y-y_0}{w_y}$ related to $x_0$ in the Hermite polynomial. This causes an overall shift in the Hermite polynomial dependent on $x_0$, which strengthens the Gaussian suppression on one end while weakening it on the other. As a result, the energy converges predominantly into a single sidelobe, as observed.

This process is illustrated in Fig. S2. In Fig. S2 (a), the blue region represents the range of influence of the Hermite polynomial without decay, while the red region shows the actual optical field distribution after balancing growth and decay. In Fig. S2 (b), depicting the case with spatial displacement, the black segment corresponds to the displacement magnitude, indicating complete suppression of a portion of one sidelobe (reducing its size). The gray area represents the attenuated intensity of this sidelobe compared to the un-displaced case, while the yellow region signifies the expanded range and enhanced intensity of the opposite sidelobe. The combined length of the yellow and gray regions matches the red segment in Fig. S2 (b).

According to angular spectrum theory, the pattern on the CCD can be treated as coherent superposition of plane waves. When no temporal displacement is introduced, the phase difference between lobes in the final spectrum remains $\pi$. However, when a temporal displacement $t_0$ is applied, the shift-phase theorem dictates that a $t_0$-dependent phase shift must be introduced in the frequency domain. This disrupts the original coherent phase relationship, ultimately causing the dark fringe between lobes to become blurred and resulting in "adhesion" between the two lobes. Mathematically, this manifests as the introduction of an imaginary term $i\frac{t_0}{w_t}$ in the Hermite polynomial.

Therefore, the singularity displacement in the frequency domain of the SLM corresponds to spatial displacement of the singularity in the spatiotemporal domain, while the singularity displacement in the spatial domain of the SLM corresponds to temporal displacement in the spatiotemporal domain.

**Mesh of Spatiotemporal Optical Vortices with Programmable Intensity Nulls Wavepackets**

Under the scalar, paraxial, slowly varying envelope, and narrow bandwidth approximations, the electric field of an optical wave packet propagating in a uniform medium with anomalous group velocity dispersion (GVD) can be expressed as the product of:

$$\frac{\partial^2 \psi}{\partial x^2} + \frac{\partial^2 \psi}{\partial y^2} + \frac{\partial^2 \psi}{\partial t^2} + 2i\frac{\partial \psi}{\partial z} = 0 \tag{S4}$$

where $\Psi$ represents the scalar wavefunction, $x$ ($x = \frac{x'}{w_x}$) and $y$ ($y = \frac{y'}{w_y}$) denote the normalized transverse coordinates, t ($t = \frac{t'}{w_t}$) is the normalized retarded time, and $z$ stands for the propagation distance. As for $x'$, $y'$ are the actual spatial coordinates and $t'$ is the actual temporal coordinates. And $w_x$, $w_y$, $w_t$ are the width of wavepackets along x direction, y direction, t direction, respectively. Given that $\Psi$ varies slowly in the $x$-direction, its first derivative $\frac{\partial^2 \psi}{\partial x^2}$ can be neglected. Anomalous group-velocity dispersion is selected to ensure that the signs of the spatial and temporal second-derivative terms coincide. To derive Eq (S4), $\frac{1}{k_0 w_y^2} = \frac{1}{w_t^2}\left|\frac{\partial^2 k}{\partial \omega^2}\right|$, where $k_0$ denotes the wave number of the central frequency, and $\frac{\partial^2 k}{\partial \omega^2}$ represents the GVD parameter evaluated at $\omega_0$.

Mapping Eq (S4) to polar coordinates, we can obtain:

$$\frac{\partial^2 \psi}{\partial r^2} + \frac{1}{r}\frac{\partial \psi}{\partial r} + \frac{1}{r^2}\frac{\partial^2 \psi}{\partial \varphi^2} + 2i\frac{\partial \psi}{\partial z} = 0 \tag{S5}$$

where $r$, $\varphi$ represent the radial coordinate and angular coordinates in the $y - t$ plane, $r = \sqrt{y^2 + t^2}$, and $\varphi = tan^{-1}\left(\frac{t}{x}\right)$.

Assume a trial solution:

$$\psi = g\left(\frac{r}{w}\right) h(r, \varphi, z) = g\left(\frac{r}{w}\right) \exp\left(-iP + \frac{i}{2q}r^2 - i\sum_{j=1}^n l_j \varphi_j\right) \quad (S6)$$

where $w$, $P$, $q$ depend on $z$ and $l_j$ take integer values. $\varphi_j = \tan^{-1}\left(\frac{(t\prime-t_j)/w_{tj}}{(x\prime-x_j)/w_{xj}}\right)$ is the function of $\varphi$. Substituting Eq (S6) to Eq (S5) reads:

$$\frac{1}{w^2} h g'' + \left(\frac{2ir}{qw} h + \frac{1}{wr} h - \frac{2ir\dot{w}}{w^2} h\right) g'$$

$$+ \left(-\frac{r^2}{q^2} h + \frac{2i}{q} h + 2P' h + \frac{r^2 q'}{q^2} h - i\frac{1}{r^2}\sum_{j=1}^n l_j \ddot{\varphi}_j h - \frac{1}{r^2}\left(\sum_{j=1}^n l_j \dot{\varphi}_j\right)^2 h\right) g = 0 \quad (S7)$$

where $h = \exp\left(-iP + \frac{i}{2q}r^2 - i\sum_{j=1}^n l_j \varphi_j\right)$ and $f = \frac{2i}{q} + 2P'$. The value of $q'$ must be 1, as required by the condition that Eq. (4) remains bounded in the limit of large $r$. Accordingly, we have the relationships:

$$\frac{1}{q} = \frac{1}{z - iz_0} = \frac{1}{R} + \frac{2i}{w^2} \quad (S8)$$

$$R' = 1 - \frac{4R^2}{w^4} \quad (S9)$$

$$w' = \frac{w}{R} \quad (S10)$$

where R is a function of z. Substituting Eq (S8), Eq (S9), Eq(S10) to Eq (S7) we can find:

$$g'' + \left(\frac{w}{r} - \frac{4r}{w}\right) g' - \left(\frac{w^2}{r^2}\left(i\sum_{j=1}^n l_j \ddot{\varphi}_j + \left(\sum_{j=1}^n l_j \dot{\varphi}_j\right)^2\right) - w^2 f\right) g = 0 \quad (S11)$$

Assuming the expression of g:

$$g = \prod_{j=1}^n 2^{\frac{l_j}{2}} \xi_j^{l_j} L(2\xi^2) \quad (S12)$$

where $\xi = \frac{r}{w}$ and $\xi_j = \frac{r_j}{w}$. $r_j = \sqrt{\left(\frac{y-y_j}{w_{yj}}\right)^2 + \left(\frac{t-t_j}{w_{tj}}\right)^2}$, $y_j$ and $t_j$ are the displacement in space and time of j-th singularity, respectively. $w_{yj}$ and $w_{tj}$ mean the spatial influence range and the temporal influence range of j-th singularity. Substituting Eq (S12) to Eq (S11) one finds:

$$2\xi^2 L'' + (\sigma + 1 - 2\xi^2) L' + \left(\Delta - \frac{\sigma}{2} + \frac{w^2 f}{8}\right) L = 0 \quad (S13)$$

where $\Delta = \frac{1}{8}\left(\sum_{j=1}^n \frac{l_j}{\xi_j} \frac{d\xi_j}{d\xi}\right)^2 - \frac{1}{8\xi^2}\left(i\sum_{j=1}^n l_j \ddot{\varphi}_j + \left(\sum_{j=1}^n l_j \dot{\varphi}_j\right)^2\right) - \frac{1}{8}\sum_{j=1}^n \frac{l_j}{\xi_j^2}\left(\frac{d\xi_j}{d\xi}\right)^2 +$

$\frac{1}{8}\frac{1}{\xi}\sum_{j=1}^n \frac{l_j}{\xi_j} \frac{d\xi_j}{d\xi} + \frac{1}{8}\sum_{j=1}^n \frac{l_j}{\xi_j} \frac{d^2\xi_j}{d\xi^2}$ and $\sigma = \xi \sum_{j=1}^n \frac{l_j}{\xi_j} \frac{d\xi_j}{d\xi}$. Set $\alpha = 2\xi^2$ and $m = \left(\Delta - \frac{\sigma}{2} + \frac{w^2 f}{8}\right)$ (m is an integer) we can obtain:

$$\alpha L'' + (\sigma + 1 - \alpha) L' + m L = 0 \quad (S14)$$

Eq (S14) is the associated Laguerre equation.

$$f = \frac{8m + 4\sigma - 8\Delta}{w^2} = \frac{2i}{q} + 2P' \quad (S15)$$

$$P' = \frac{z}{i(z^2 + z_0^2)} + \frac{z_0(2m + n + 1 - 2\Delta)}{z^2 + z_0^2} \quad (S16)$$

$$exp(-iP) = \left(1 + \frac{z^2}{z_0^2}\right)^{\frac{-1}{2}} exp\left(-i(2m + \sigma + 1 - 2\Delta) tan^{-1}\left(\frac{z}{z_0}\right)\right) \tag{S18}$$

Consequently, the scalar wave function can be expressed as:

$$\frac{w_0}{w} \prod_{j=1}^{n} \left(\frac{\sqrt{2}r_j}{w}\right)^{l_j} L_m^\sigma\left(\frac{2r^2}{w^2}\right) exp\left(-\frac{r^2}{w^2}\right) exp\left(\frac{ir^2}{2R} - i\sum_{j=1}^{n} l_j \varphi_j - i(2m + \sigma + 1 - 2\Delta) tan^{-1}\left(\frac{z}{z_0}\right)\right) \tag{S19}$$

where $R = z\left(1 + \left(\frac{z_0}{z}\right)^2\right)$, $w = w_0\sqrt{1 + \left(\frac{z}{z_0}\right)^2}$, $w_0 = \sqrt{2z_0}$, $L_m^\sigma(\cdot)$ is a generalized Laguerre polynomial, as $m$ and $\sigma$ are the radial and angular mode numbers. At $z = 0$, the wave function reads:

$$\psi_{z=0} = \prod_{j=1}^{n} \left(\frac{\sqrt{2}r_j}{w}\right)^{l_j} L_m^\sigma\left(\frac{2r^2}{w^2}\right) exp\left(-\frac{r^2}{w^2}\right) exp\left(-i\sum_{j=1}^{n} l_j \varphi_j\right) \tag{S20}$$

**Spectrum of M-STOV:**

According to the analysis of the diffraction detection of M-STOV, their spectrum can be calculated as:

$$\mathcal{F}_t[\psi(y,t)] = f_{j,j\geq 2}(y,\omega) * g(y,\omega) \tag{S21}$$

We begin by calculating $f_{j,j\geq 2}(y,\omega)$

$$f_{j,j\geq 2}(y,\omega) = \int_{-\infty}^{\infty} \left(\frac{t-t_j}{w_{tj}} + isgn(l_i)\frac{y}{w_{yj}}\right)^{|l_j|} e^{-i\omega t} dt \tag{S22}$$

According to the binomial theory (39), Eq (S22) can be rewritten as:

$$\sum_{k=0}^{|l_j|} \binom{|l_j|}{k} i^{|l_j|-k} \left(sgn(l_j)\frac{y}{w_{yj}}\right)^{|l_i|-k} \int_{-\infty}^{\infty} \left(\frac{t-t_j}{w_{tj}}\right)^k e^{-i\omega t} dt \tag{S23}$$

Set $u = \frac{t-t_j}{w_{tj}}$, Eq (S23) reads:

$$\sum_{k=0}^{|l_j|} \binom{|l_j|}{k} i^{|l_j|-k} \left(sgn(l_j)\frac{y}{w_{yj}}\right)^{|l_i|-k} \int_{-\infty}^{\infty} w_{tj}(u)^k e^{-i\omega(w_{tj}u+t_j)} du \tag{S24}$$

Calculating Eq (S24), we can find:

$$f_{j,j\geq 2}(y,\omega) = 2\pi i^{|l_j|} \sum_{k=0}^{|l_j|} \binom{|l_j|}{k} \left(sgn(l_j)\frac{y}{w_{yj}}\right)^{|l_j|-k} e^{-i\omega t_j} \frac{1}{w_{tj}^k} \delta^{(k)}(\omega) \tag{S25}$$

We now proceed to calculate $g(y,\omega)$:

$$g(y,\omega) = \int_{-\infty}^{\infty} \left(\frac{t-t_0}{w_{t1}} + isgn(l)\frac{y}{w_{y1}}\right)^{|l|} exp\left(-\frac{t^2}{w_{t2}^2} - \frac{x^2}{w_{x2}^2}\right) e^{-i\omega t} dt$$

$$= exp\left(-\frac{y^2}{w_{y2}^2}\right) \int_{-\infty}^{\infty} \left(\frac{t-t_0}{w_{t1}} + isgn(l)\frac{y}{w_{y1}}\right)^{|l|} exp\left(-\frac{t^2}{w_{t2}^2}\right) e^{-i\omega t} dt \tag{S26}$$

Set:

$$J = \int_{-\infty}^{\infty} \left(\sqrt{c_1}t - \sqrt{c_1}t_0 + isgn(l)\frac{y}{w_{y1}}\right)^{|l|} exp(-c_2 t^2) e^{-i\omega t} dt \tag{S27}$$

where $c_1 = \frac{1}{w_{t1}^2}$ and $c_2 = \frac{1}{w_{t2}^2}$

According to the binomial theory, Eq (S27) reads:

$$J = \sum_{k=0}^{|l|} \binom{|l|}{k} \left(isgn(l)\frac{y}{w_{y1}} - \sqrt{c_1}t_0\right)^k \int_{-\infty}^{\infty} (\sqrt{c_1}t)^{|l|-k} exp(-c_2 t^2) e^{-i\omega t} dt \tag{S28}$$

Set $b = -i\omega$ and $u = t - \frac{b}{2c_2}$, and employ the converse of the binomial theorem:

$$J = e^{\frac{-\omega^2}{4c_2}} \int_{-\infty}^{\infty} (\sqrt{c_1})^{|l|} \left( u - \frac{i\omega}{2c_2} + i\,sgn(l)\frac{y}{w_{y1}\sqrt{c_1}} - t_0 \right)^{|l|} e^{-c_2 u^2} du \tag{S29}$$

We can assume that:

$$g(T) = \int_{-\infty}^{\infty} \exp[T(u+ia)] e^{-c_2 u^2} du \tag{S30}$$

where $a = -\frac{\omega}{2c_2} + sgn(l)\frac{y}{w_{y1}\sqrt{c_1}} + it_0$.

Notice that when $T = 0$, we can obtain:

$$\frac{d^{|l|}g(T)}{dT^{|l|}} = \int_{-\infty}^{\infty} (u+ia)^{|l|} e^{-c_2 u^2} du \tag{S31}$$

Calculating $g(T)$, we can find:

$$g(T) = \sqrt{\frac{\pi}{c}} e^{c_2 a^2} \exp\left[\frac{1}{4c_2}(T+2iac_2)^2\right] \tag{S32}$$

Set $s = T + 2iac_2$ and $k = \frac{1}{4c_2}$, we have:

$$\frac{d^{|l|}g(t)}{dT^{|l|}} = \sqrt{\frac{\pi}{c}} e^{c_2 a^2} \frac{d^{|l|}}{ds^{|l|}} \exp[ks^2] \tag{S33}$$

Set $v = \sqrt{k}s$, we have:

$$\frac{d^{|l|}}{ds^{|l|}} \exp[ks^2] = k^{\frac{|l|}{2}} \frac{d^{|l|}}{dv^{|l|}} \exp(v^2) \tag{S34}$$

The definition of Hermite polynomial is:

$$H_n(x) = (-1)^n e^{x^2} \frac{d^n}{dx^n} e^{-x^2} \tag{S35}$$

$$H_n(ix) = (i)^n e^{-x^2} \frac{d^n}{dx^n} e^{x^2} \tag{S36}$$

Hence:

$$k^{\frac{|l|}{2}} \frac{d^{|l|}}{dv^{|l|}} \exp(v^2) = k^{\frac{|l|}{2}} (-i)^{|l|} H_{|l|}(i\sqrt{k}s) e^{ks^2} \tag{S37}$$

Accordingly, we can obtain:

$$\frac{d^{|l|}g(T)}{dT^{|l|}} = \sqrt{\frac{\pi}{c}} e^{c_2 a^2} k^{\frac{|l|}{2}} (-i)^{|l|} H_{|l|}\left( \frac{i}{2\sqrt{c_2}}(T+2iac_2) \right) e^{\frac{1}{4c_2}(T+2iac_2)^2} \tag{S38}$$

When $T = 0$, we can find:

$$\sqrt{\frac{\pi}{c}} e^{c_2 a^2} k^{\frac{|l|}{2}} (-i)^{|l|} H_{|l|}\left( \frac{i}{2\sqrt{c_2}}(T+2iac_2) \right) e^{\frac{1}{4c_2}(T+2iac_2)^2} = \sqrt{\frac{\pi}{c_2}} i^{|l|} (4c_2)^{\frac{-|l|}{2}} H_{|l|}(\sqrt{c_2}a) \tag{S39}$$

Therefore, $J$ can be expressed as:

$$J = e^{\frac{-\omega^2}{4c_2}} (\sqrt{c_1})^{|l|} \sqrt{\frac{\pi}{c_2}} i^{|l|} (4c_2)^{\frac{-|l|}{2}} H_{|l|}(\sqrt{c_2}a) \tag{S40}$$

Substituting $c_1 = \frac{1}{w_{t1}^2}$, $c_2 = \frac{1}{w_{t2}^2}$ and $a = -\frac{\omega}{2c_2} + sgn(l)\frac{y}{w_{y1}\sqrt{c_1}} + it_0$ to Eq (S40), we can have:

$$J = \exp\left(\frac{-w_{t2}^2 \omega^2}{4}\right) \left(\frac{1}{w_{t1}}\right)^{|l|} i^{|l|} \sqrt{\pi} w_{t2}^{|l|+1} 2^{-|l|} H_{|l|}\left( \frac{1}{w_{t2}}\left(sgn(l)\frac{yw_{t1}}{w_{y1}} - \frac{\omega w_{t2}^2}{2} + it_0 \right) \right) \tag{S41}$$

Thereby, $g(y,\omega)$ can be expressed as:

$$g = \left(\frac{1}{w_{t1}}\right)^{|l|} i^{|l|} \sqrt{\pi} w_{t2}^{|l|+1} 2^{-|l|} H_{|l|}\left( sgn(l)\frac{yw_{t1}}{w_{y1}w_{t2}} - \frac{\omega w_{t2}}{2} + i\frac{t_0}{w_{t2}} \right) \exp\left(-\frac{y^2}{w_{y2}^2} - \frac{w_{t2}^2 \omega^2}{4}\right) \tag{S42}$$

**The spectral analytical expressions for two singularities M-STOV:**

According to Eq (3) and Eq (4), the spectrum of two singularities M-STOV can be calculated as:

$$\mathcal{F}_t\left[\left(\frac{t-t_1}{w_{t1}} + i\,sgn(l_1)\frac{y-y_1}{w_{y1}}\right)^{|l_1|}\right] * \mathcal{F}_t\left[\left(\frac{t-t_2}{w_{t2}} + i\,sgn(l_2)\frac{y-y_2}{w_{y2}}\right)^{|l_2|} exp\left(-\frac{y^2}{w_{y3}^2} - \frac{t^2}{w_{t3}^2}\right)\right] \quad (S43)$$

Substituting Eq (S25) and Eq (S42) to Eq (S43), we can find:

$$f(\tau) = 2\pi i^{|l_1|} \sum_{k=0}^{|l_1|} \binom{|l_1|}{k} \left(sgn(l_1)\frac{y}{w_{y1}}\right)^{|l_1|-k} e^{-i\tau t_1} \frac{1}{w_{t1}^k} \delta^{(k)}(\tau) \quad (S44)$$

$$g(\tau) = \left(\frac{1}{w_{t2}}\right)^{|l_2|} i^{|l_2|} \sqrt{\pi} w_{t3}^{|l_2|+1} 2^{-|l_2|} H_{|l\_2|}\left(sgn(l_2)\frac{y w_{t2}}{w_{y2}w_{t3}} - \frac{\tau w_{t3}}{2} + i\frac{t_0}{w_{t3}}\right) exp\left(-\frac{y^2}{w_{y3}^2} - \frac{w_{t3}^2 \tau^2}{4}\right) \quad (S45)$$

$$(f * g)(\omega) = 2\pi i^{|l_1|} \sum_{k=0}^{|l_1|} \binom{|l_1|}{k} \left(sgn(l_1)\frac{y}{w_{y1}}\right)^{|l_1|-k} \frac{1}{w_{t1}^k} (-1)^k \frac{d^k}{d\tau^k}\left[e^{-i\tau t_1} g(\omega - \tau)\right]_{\tau=0} \quad (S46)$$

Set:

$$a = sgn(l_2)\frac{y w_{t2}}{w_{y2}w_{t3}} + i\frac{t_0}{w_{t3}} \quad (S47)$$

$$b = \frac{w_{t3}}{2} \quad (S48)$$

$$c = \left(\frac{1}{w_{t2}}\right)^{|l_2|} i^{|l_2|} \sqrt{\pi} w_{t3}^{|l_2|+1} 2^{-|l_2|} exp\left(-\frac{y^2}{w_{y3}^2}\right) \quad (S49)$$

$$h(\tau) = e^{-i\tau t_1} g(\omega - \tau) \quad (S50)$$

Hence:

$$g(\omega - \tau) = c H_{|l\_2|}[a - b(\omega - \tau)] exp\left(-\frac{w_{t3}^2(\omega-\tau)^2}{4}\right) \quad (S51)$$

And set:

$$u(\tau) = e^{-i\tau t_1} \quad (S52)$$
$$v(\tau) = H_{|l_2|}[a - b(\omega - \tau)] \quad (S53)$$
$$w(\tau) = exp\left(-\frac{w_{t3}^2(\omega-\tau)^2}{4}\right) \quad (S54)$$

We can get:

$$h(\tau) = e^{-i\tau t_1} g(\omega - \tau) = c u(\tau) v(\tau) w(\tau) \quad (S55)$$

According to the Leibniz's rule for differentiation (*40*):

$$\frac{d^k}{d\tau^k}\left[e^{-i\tau t_1} g(\omega - \tau)\right]_{\tau=0} = \sum_{n,j,m\geq 0; i+j+m=k} \binom{k}{n,j,m} u^n(0) v^j(0) w^m(0) \quad (S56)$$

Calculating $u^n(0)$, we have:

$$u^n(0) = \frac{d^n}{d\tau^n} e^{-i\tau t_1} = (-it_1)^n \quad (S57)$$

Calculating $v^j(0)$, we have:

$$v^j(0) = (w_{t3})^j \frac{|l_2|!}{(|l_2|-j)!} H_{|l\_2|-j}\left(sgn(l_2)\frac{y w_{t2}}{w_{y2}w_{t3}} + i\frac{t_0}{w_{t3}} - \frac{w_{t3}}{2}\omega\right) \quad (S58)$$

And for $j > |l_2|$, $v^j(0) = 0$.

Calculating $w^m(0)$, we have:

$$w^m(0) = \left(\frac{w_{t3}}{2}\right)^m exp\left(-\frac{w_{t3}^2}{4}(\omega)^2\right) H_m\left(\frac{w_{t3}}{2}\omega\right) \quad (S59)$$

Substituting Eq (S55)-Eq (S59) to Eq (S46), we can obtain:

$$(f * g)(\omega) = 2\pi i^{|l_1|+|l_2|} \left(\frac{1}{w_{t2}}\right)^{|l_2|} \sqrt{\pi} w_{t3}^{|l_2|+1} 2^{-|l_2|} \exp\left(-\frac{x^2}{w_{x3}^2} - \frac{w_{t3}^2 \omega^2}{4}\right) \times$$

$$\sum_{k=0}^{|l_1|} \binom{|l_1|}{k} \left(sgn(l_1)\frac{x}{w_{x1}}\right)^{|l_1|-k} \frac{1}{w_{t1}^k}(-1)^k$$

$$\left\{ \begin{array}{c} \sum_{n,j,m\geq 0; i+j+m=k} \binom{k}{n,j,m}(-it_1)^n (w_{t3})^j \frac{|l_2|!}{(|l_2|-j)!} \\ H_{|l\_2|-j}\left(sgn(l_2)\frac{xw_{t2}}{w_{x2}w_{t3}} + i\frac{t_0}{w_{t3}} - \frac{w_{t3}}{2}\omega\right)\left(\frac{w_{t3}}{2}\right)^m H_m\left(\frac{w_{t3}}{2}\omega\right) \end{array} \right\} \quad (S60)$$

**Additional Combination Cases of diffraction patterns of M-STOV**

Fig. S3 shows the combination case with:

$$\begin{bmatrix} 0 & 1 \\ 0 & 1 \end{bmatrix} \boxplus \begin{bmatrix} 0 & 0 \\ -1 & 0 \end{bmatrix} = \begin{bmatrix} 0 & 1 \\ -1 & 1 \end{bmatrix}$$

The overall shape of $M = \begin{bmatrix} 0 & 1 \\ -1 & 1 \end{bmatrix}$ structure is primarily comprised of the detection pattern for element $M_1 = \begin{bmatrix} 0 & 1 \\ 0 & 1 \end{bmatrix}$ (The first column), while the form of its right-side is modulated by the diffraction pattern of element $M_2 = \begin{bmatrix} 0 & 0 \\ -1 & 0 \end{bmatrix}$ (The second column). The superposition region of elements $M_1 = \begin{bmatrix} 0 & 1 \\ 0 & 1 \end{bmatrix}$ and $M_2 = \begin{bmatrix} 0 & 0 \\ -1 & 0 \end{bmatrix}$ is smoothed by the convolution process, resulting in the distortion of certain distinct. The white dotted line in the figure divides the approximate area influenced by $M_1$ and $M_2$.

Fig. S4 shows the combination case with:

$$\begin{bmatrix} -1 & 1 \\ -1 & 0 \end{bmatrix} \boxplus \begin{bmatrix} 0 & 0 \\ 0 & 1 \end{bmatrix} = \begin{bmatrix} -1 & 1 \\ -1 & 1 \end{bmatrix}$$

The overall shape of $M = \begin{bmatrix} -1 & 1 \\ -1 & 1 \end{bmatrix}$ structure is primarily comprised of the detection pattern for element $M_1 = \begin{bmatrix} -1 & 1 \\ -1 & 0 \end{bmatrix}$ (The first column), while the form of its upper-right side is modulated by the diffraction pattern of element $M_2 = \begin{bmatrix} 0 & 0 \\ 0 & 1 \end{bmatrix}$ (The second column). The superposition region of elements $M_1 = \begin{bmatrix} -1 & 1 \\ -1 & 0 \end{bmatrix}$ and $M_2 = \begin{bmatrix} 0 & 0 \\ 0 & 1 \end{bmatrix}$ is smoothed by the convolution process, resulting in the distortion of certain distinct. The white dotted line in the figure divides the approximate area influenced by $M_1$ and $M_2$.

**Detection of M-STOV Second-Harmonic Generation Field**

Our SHG results for the M-STOVs-PINs are presented in Fig. S5, where the selected parameters are $M = \begin{bmatrix} 0 & 0 \\ -1 & 0 \end{bmatrix}, \begin{bmatrix} -1 & 0 \\ 0 & -1 \end{bmatrix}, \begin{bmatrix} -1 & 0 \\ -1 & 1 \end{bmatrix}$ and $\begin{bmatrix} 1 & -1 \\ 1 & -1 \end{bmatrix}$. The first row of Fig. S5 shows the detection pattern on CCD for 800 nm field, and the second row illustrates the space-spectra of the second-harmonic field. The doubling of topological charge during the SHG process can be distinctly observed across these scenarios. The white dashed line in the figure separates regions with opposite topological charge helicity.

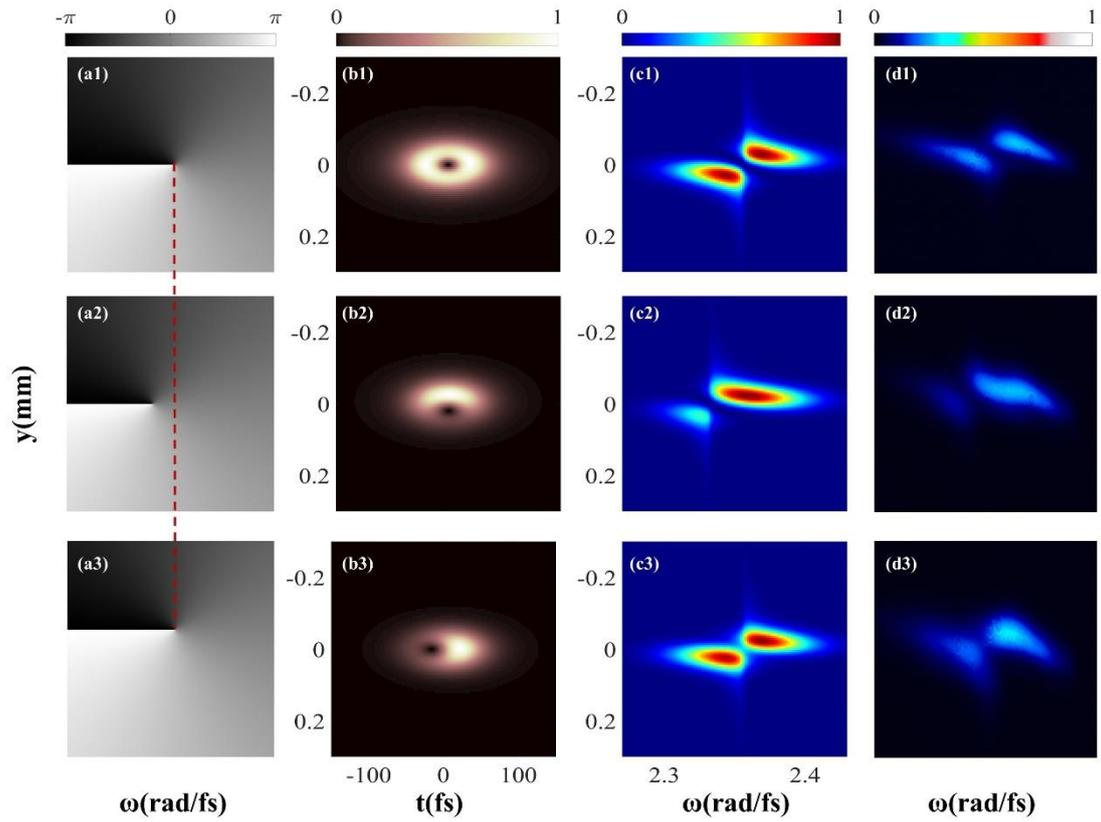

**Fig. S1. Experimental and simulated results of first-order STOV generated by shifting the phase singularity position in both SLM frequency and spatial domains.** Column 1 (**a1-a3**) presents the phase patterns loaded on SLM. Column 2 (**b1-b3**) shows the corresponding simulated spatiotemporal intensity distribution. Column 3 (**c1-c3**) displays the simulated diffraction patterns. Column 4 (**d1-d3**). contains the measured diffraction patterns.

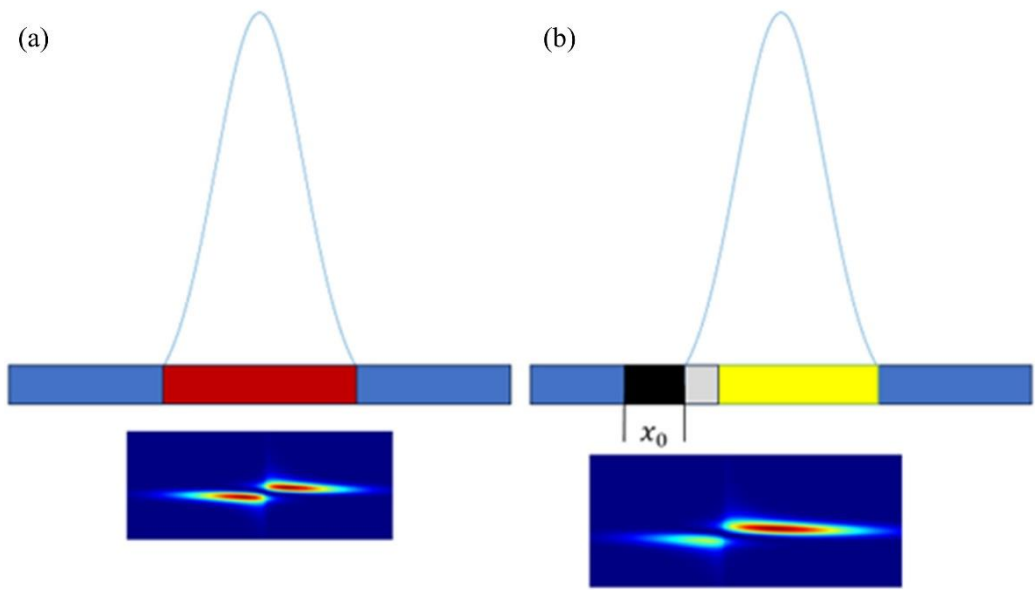

**Fig. S2. Impact of frequency-dimensional singularity displacement in SLM on the detected diffraction pattern. (a)** the blue region represents the range of influence of the Hermite polynomial without decay, while the red region shows the actual optical field distribution after balancing growth and decay. **(b)** the black segment corresponds to the displacement magnitude, indicating complete suppression of a portion of one sidelobe (reducing its size). The gray area represents the attenuated intensity of this sidelobe compared to the un-displaced case, while the yellow region signifies the expanded range and enhanced intensity of the opposite sidelobe. The blue bell-shaped curve in this figure represents a Gaussian distribution.

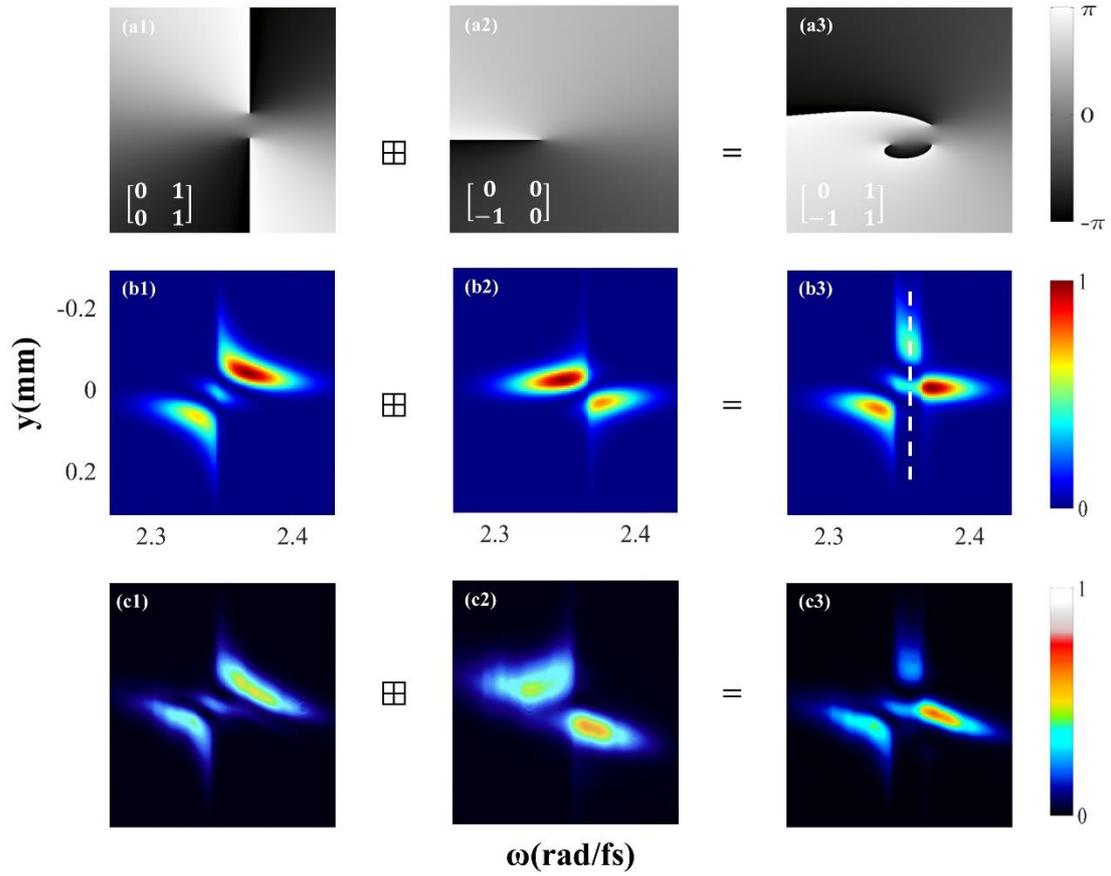

**Fig. S3. Simulated and experiment results of M-STOV with topological charges arrangement** $\begin{bmatrix} 0 & 1 \\ 0 & 1 \end{bmatrix} \boxplus \begin{bmatrix} 0 & 0 \\ -1 & 0 \end{bmatrix} = \begin{bmatrix} 0 & 1 \\ -1 & 1 \end{bmatrix}$. Row 1 (**a1–a3**) displays the phase patterns loaded on SLM. Row 2 (**b1–b3**) illustrates the simulated measurement results. Row 3 (**c1–c3**) contains the corresponding experimental space-spectra. The boxed plus symbol (⊞) is introduced here to distinguish this specific process from ordinary superposition. The white text indicates the mesh of topological charges for the simulation and experiments.

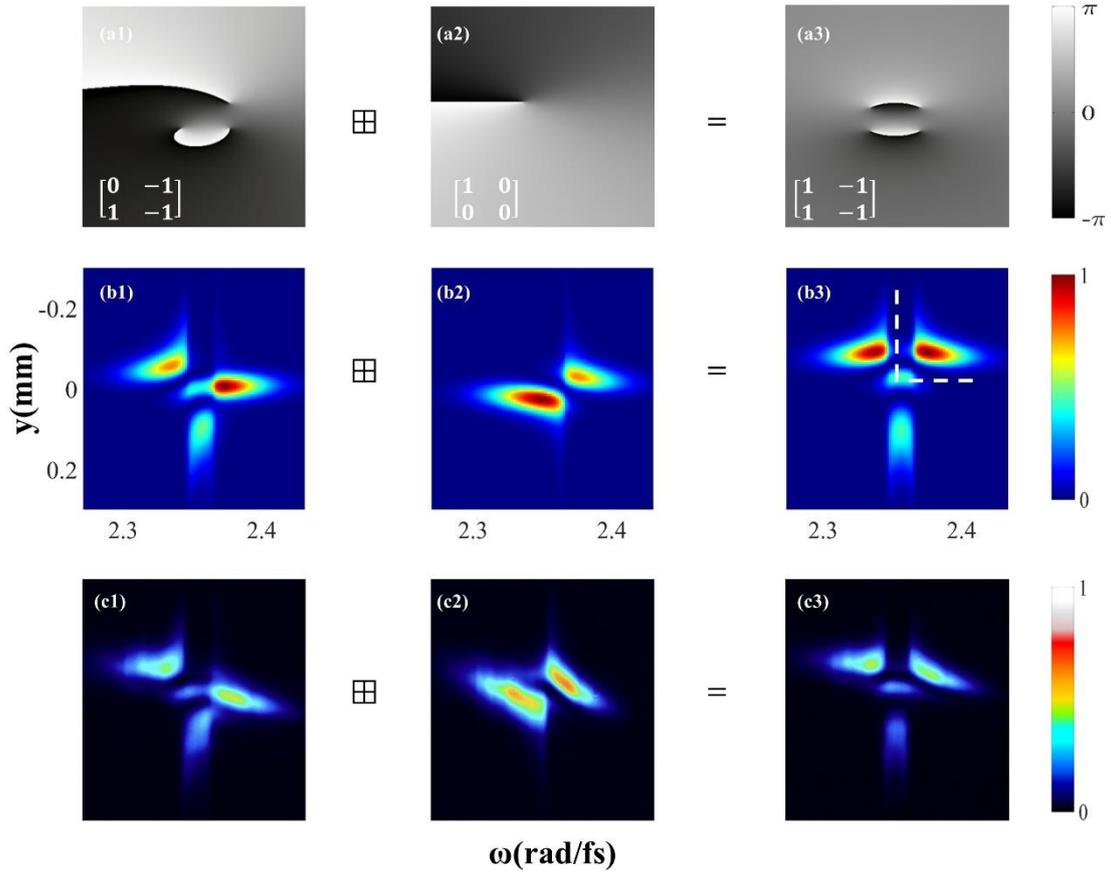

**Fig. S4. Simulated and experiment results of M-STOV with topological charges arrangement $\begin{bmatrix} 0 & -1 \\ 1 & -1 \end{bmatrix} \boxplus \begin{bmatrix} 1 & 0 \\ 0 & 0 \end{bmatrix} = \begin{bmatrix} 1 & -1 \\ 1 & -1 \end{bmatrix}$.** Row 1 (**a1–a3**) displays the phase patterns loaded on SLM. Row 2 (**b1–b3**) illustrates the simulated measurement results. Row 3 (**c1–c3**) contains the corresponding experimental space-spectra. The boxed plus symbol (⊞) is introduced here to distinguish this specific process from ordinary superposition. The white text indicates the mesh of topological charges for the simulation and experiments.

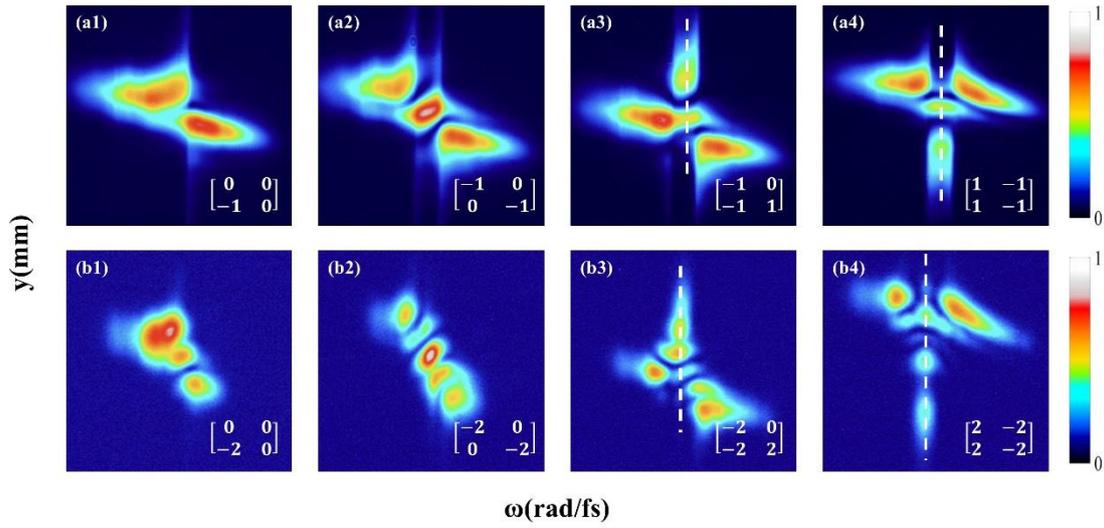

**Fig. S5. Measured space-spectrum of M-STOV with topological charges arrangement** $M = \begin{bmatrix} 0 & 0 \\ -1 & 0 \end{bmatrix}, \begin{bmatrix} -1 & 0 \\ 0 & -1 \end{bmatrix}, \begin{bmatrix} -1 & 0 \\ -1 & 1 \end{bmatrix}$ **and** $\begin{bmatrix} 1 & -1 \\ 1 & -1 \end{bmatrix}$ **at 800nm and 400nm.** Row1 (**a1-a4**) shows the detection pattern on CCD for 800 nm field. Row2 (**b1-b4**) illustrates the space-spectra of the second-harmonic field. The white dashed line in the figure separates regions with opposite topological charge helicity.